\documentclass[doublecol,linenumbers]{epl2} 

\usepackage{epstopdf}
\usepackage{color}
\usepackage{subfigure}
\usepackage{stfloats}
\usepackage{amsmath}
\usepackage{amssymb}
\usepackage{graphicx,color}
\usepackage{cite}
\usepackage{enumerate}
\usepackage{amsthm}
\usepackage{amsfonts,mathrsfs}
\usepackage{geometry}
\usepackage{ifthen}

\title{\bf Uncertainty and complementarity relations based on generalized skew information}
\shorttitle{Uncertainty and complementarity relations based on generalized skew information} 

\author{Huaijing Huang\inst{1} \and Zhaoqi Wu\inst{1}\thanks{E-mail: wuzhaoqi\_conquer@163.com (corresponding author)} \and  Shao-Ming
Fei\inst{2,3}} \shortauthor{Huaijing Huang  \etal}

\institute{
  \inst{1} Department of Mathematics, Nanchang University,
Nanchang 330031, P R China\\
  \inst{2} School of Mathematical Sciences, Capital Normal University, Beijing 100048,
  China\\
  \inst{3} Max Planck Institute for Mathematics in the Sciences, 04103 Leipzig, Germany
}

\abstract{ Uncertainty relations and complementarity relations are
core issues in quantum mechanics and quantum information theory. By
use of the generalized Wigner-Yanase-Dyson (GWYD) skew information,
we derive several uncertainty and complementarity relations with
respect to mutually unbiased measurements (MUMs), and general
symmetric informationally complete positive operator valued
measurements (SIC-POVMs), respectively. Our results include some
existing ones as particular cases. We also exemplify our results by
providing a detailed example.}

\begin{document}

\maketitle

\section{\bf  Introduction }
As embodiment of the Heisenberg uncertainty principle, uncertainty
relations form a central part of our understanding on quantum
mechanics, providing fundamental constraints on how well the
outcomes of various incompatible measurements can be predicted.
Heisenberg first noted the uncertainty in the measurements of
position and momentum \cite{HW}. Robertson further generalized it to
two arbitrary observables and presented a lower bound on the total
variance of two observables \cite{RH}. Uncertainty relations are
generally referred to as the lower bounds on the quantifiers.
Various quantitative characterizations including entropy
\cite{DD,HM,AE,FG}, Wigner-Yanase skew information
\cite{BC,SL1,SS,ZW,MCF,LF}, variance \cite{CB1} and statistical
distance \cite{QHH} have been extensively studied.

Quantum measurement plays fundamental roles in quantum mechanics.
Different kinds of measurements including von Neumann measurements
\cite{JV}, L\"uders measurements \cite{GL}, the dissipative
adiabatic measurements (DAMs) \cite{ZD}, symmetric informationally
complete positive operator valued measures (SIC-POVMs) \cite{RJ},
general SIC-POVMs \cite{AD} have been introduced and investigated.
Since quantum coherence is basis-dependent, it is natural to study
uncertainty relations of coherence with respect to a given
measurement basis.

As the most basic feature in quantum mechanics, quantum coherence is
extremely significant physical resource. The problem of properly
quantifying coherence at the quantum level has attracted
considerable attention, there are many different ways to measure
coherence\cite{ZD2,BT2,SA2,XJ2}. In \cite{SL3}, the author revealed
the close relationship between coherence and the quantum part of
uncertainties. The concept of quantum uncertainty relations of
quantum coherence has been introduced in \cite{US,XY}. By deriving
the upper bounds on the sum of the corresponding measures, the
complementarity relations of quantum coherence in different bases
have been studied \cite{SC,FP}. In addition, by using the
Wigner-Yanase (WY) skew information, Luo et al. \cite{SL7} not only
studied a quantitative complementarity relation in the ubiquitous
state-channel interaction, but also extended the coherence of $\rho$
with respect to an orthonormal basis to the one with respect to a
quantum channel $\Phi$. Furthermore, Wu et al. \cite{WZ} discussed
the complementarity relations and coherence measures by using the
modified GWYD skew information. Uncertainty relations for quantum
coherence with respect to mutually unbiased bases (MUBs) has also
been investigated \cite{RA}.

In terms of the coherence measure based on the WY skew information,
several uncertainty relations for coherence with respect to von
Neumann measurements, MUBs and general SIC-POVMs have been
established, respectively \cite{SL2}. The average coherence of a
state with respect to any complete set of mutually unbiased
measurements (MUMs) and general SIC-POVMs has been also evaluated,
respectively \cite{CB2}. A natural question arises: can we consider
the uncertainty and complementarity relations for measures based on
GWYD skew information with respect to MUMs and general SIC-POVMs? We
study these problems in this paper.

\section{\bf  GWYD skew information
and quantum uncertainty } Let $\mathcal{H}$ be a $d$-dimensional
Hilbert space, and $\mathcal{S(H)}$ and $\mathcal{D(H)}$ the set of
Hermitian operators and density operators on $\mathcal{H}$,
respectively. For a density operator $\rho\in \mathcal{D(H)}$ and an
observable $A\in \mathcal{S(H)}$, the WY skew information \cite{WY}
is defined by
\begin{equation}\label{eq1}
I_{\rho}(A)=-\frac{1}{2}\mathrm{Tr}([\rho^{\frac{1}{2}},A]^2),
\end{equation}
where $[X,Y]:=XY-YX$ is the commutator of $X$ and $Y$. A more
general quantity was proposed by Dyson,
\begin{equation}\label{eq2}
I_{\rho}^{\alpha}(A)=-\frac{1}{2}\mathrm{Tr}([\rho^{\alpha},A][\rho^{1-\alpha},A]),~0\leq
\alpha \leq 1,
\end{equation}
which is now called the  Wigner-Yanase-Dyson (WYD) skew information.
The quantity in Eq.~(\ref{eq2}) was further generalized to \cite{CL}
\begin{equation}\label{eq3}
I_{\rho}^{\alpha,\beta}(A)=-\frac{1}{2}\mathrm{Tr}([\rho^\alpha,
A][\rho^\beta, A]\rho^{1-\alpha-\beta}),
\end{equation}
with $\alpha,\beta\geq 0,\alpha+\beta\leq 1$, which is termed as
GWYD skew information. It is easy to see that when $\alpha+\beta=1$,
$I_{\rho}^{\alpha,\beta}(A)$ reduces to $I_{\rho}^{\alpha}(A)$, and
$I_{\rho}^{\alpha}(A)$ reduces to $I_{\rho}(A)$ when
$\alpha=\frac{1}{2}$. $I_{\rho}^{\alpha,\beta}(A)$ can be
equivalently expressed as
\begin{equation}\label{eq4}
\aligned I_{\rho}^{\alpha,\beta}(A)&=\frac{1}{2}[\mathrm{Tr}(\rho
A^2)+\mathrm{Tr}(\rho^{\alpha+\beta} A\rho^{1-\alpha-\beta}A)\\
&\quad-\mathrm{Tr}(\rho^{\alpha}
A\rho^{1-\alpha}A)-\mathrm{Tr}(\rho^{\beta} A\rho^{1-\beta}A)],
\endaligned
\end{equation}
where $\alpha,\beta\geq 0$ and $\alpha+\beta\leq 1$.

The set of all observables on $\mathcal{H}$ constitutes a real $d^2$
-dimensional Hilbert space $M$ with inner product $\langle
A,B\rangle =\text{Tr}AB$. Let $\{K_i\}^{d^2}_{i=1}$ be any complete
orthonormal base of $M$. In Ref. \cite{SL6} the quantum uncertainty
of a mixed state $\rho$ is defined as
\begin{equation}\label{eq5}
Q(\rho)=\sum _{i=1}^{d^2}I_{\rho}(K_i).
\end{equation}
Denote $\{\lambda_i\}^d_{i=1}$ the spectrum of $\rho$. One has
\begin{equation}\label{eq6}
 Q(\rho)=\sum_{i<j}(\sqrt{\lambda_i}-\sqrt{\lambda_j})^2=d-(\mathrm{Tr}\sqrt{\rho})^2.
\end{equation}
With respect to the WYD skew information, Li et al. \cite{LX}
proposed the quantum uncertainty of a mixed state $\rho$ as
\begin{equation}\label{eq7}
Q_{\alpha}(\rho)=\sum _{i=1}^{d^2}I_{\rho}^{\alpha}(K_i),~~~ 0\leq
\alpha \leq 1,
\end{equation}
which can be further expressed as
\begin{equation}\label{eq8}
\aligned
Q_{\alpha}(\rho)&=\sum_{i<j}(\lambda_i^\alpha-\lambda_j^\alpha)(\lambda_i^{1-\alpha}
-\lambda_j^{1-\alpha})\\&=d-\mathrm{Tr}{\rho^{\alpha}}
\mathrm{Tr}{\rho^{1-\alpha}}.
\endaligned
\end{equation}
It can be proved that $Q_{\alpha}(\rho)\leq Q(\rho)$ for $0\leq
\alpha \leq 1$. Similarly, by using the GWYD skew information define
in Eq.~(\ref{eq3}), we define the following generalized quantum
uncertainty,
\begin{equation}\label{eq9}
Q^{\alpha,\beta}({\rho})=\sum
_{i=1}^{d^2}I_{\rho}^{\alpha,\beta}(K_i),~~~\alpha,\beta\geq
0,~\alpha+\beta\leq 1,
\end{equation}
which has the form in terms of the spectrum of $\rho$,
\begin{equation}\label{eq10}
\aligned Q^{\alpha,\beta}({\rho})&=\frac{1}{2}\sum
_{i<j}[(\lambda_i^\alpha-\lambda_j^\alpha)\\&\quad\times(\lambda_i^\beta
-\lambda_j^\beta)(\lambda_i^{1-\alpha-\beta}+\lambda_j^{1-\alpha-\beta})].
\endaligned
\end{equation}
It can be seen that $Q^{\alpha,\beta}({\rho})$ reduces to
$Q_{\alpha}(\rho)$ when $\alpha+\beta=1$, and $Q_{\alpha}(\rho)$
reduces to $Q(\rho)$ when $\alpha={1}/{2}$.

{\it Remark} It should be noted that the quantity defined in
Eq.~(\ref{eq9}) is different from the one defined in Ref. \cite{CL},
in which the quantum extensions of the Fisher information has been
investigated and the quantity $Q_{\alpha,\beta}(\rho)=\sum
_{i=1}^{d^2}I_{\alpha,\beta}(\rho,K_i)$ has been defined as a
measure of the quantum uncertainty, where
$I_{\alpha,\beta}(\rho,K_i)=\frac{1}{\alpha\beta}[\mathrm{Tr}(\rho
K_i^2)+\mathrm{Tr}(\rho^{\alpha+\beta}
K_i\rho^{1-\alpha-\beta}K_i)-\mathrm{Tr}(\rho^{\alpha} K_i
\rho^{1-\alpha}K_i)-\mathrm{Tr}(\rho^{\beta} K_i\rho^{1-\beta}K_i)]$
for $\alpha,\beta\geq 0$, $\alpha+\beta\leq 1$. Further calculations
show that $Q_{\alpha,\beta}(\rho)=\frac{1}{2\alpha\beta}\sum
_{i,j=1}^{d}[(\lambda_i^\alpha-\lambda_j^\alpha)(\lambda_i^\beta
-\lambda_j^\beta)(\lambda_i^{1-\alpha-\beta}+\lambda_j^{1-\alpha-\beta})]$.

By using the inequality
$\lambda_i^{\alpha}\lambda_j^{1-\alpha}+\lambda_i^{1-\alpha}\lambda_j^{\alpha}
\geq\lambda_i^{\alpha+\beta}\lambda_j^{1-\alpha-\beta}+\lambda_i^{1-\alpha-\beta}\lambda_j^{\alpha+\beta}$
for $\alpha, \beta\in[0,1]$ with $\alpha+2\beta\leq1$ and
$2\alpha+\beta\leq1$ and Eq.~(\ref{eq10}), we can prove the
following lemma.

{\bf Lemma 1}  $Q^{\alpha,\beta}({\rho})$ satisfies the following
inequality
\begin{equation}\label{eq11}
Q^{\alpha,\beta}({\rho})\leq\frac{1}{2}(
d-\mathrm{Tr}{\rho^{\alpha}} \mathrm{Tr}{\rho^{1-\alpha}})
\end{equation}
for $\alpha,\beta\in[0,1]$ with $\alpha+2\beta\leq1$ and
$2\alpha+\beta\leq1$.

\section{\bf  Uncertainty and complementarity relations based on
generalized skew information with respect to MUMs } In this section,
we consider the uncertainty and complementarity relations based on
generalized skew information with respect to mutually unbiased
measurements (MUMs). Two orthonormal bases
$\mathcal{B}_1=\{|b_{1k}\rangle\}^d_{k=1}$ and
$\mathcal{B}_2=\{|b_{2k}\rangle\}^d_{k=1}$ of $\mathcal{H}$ are
called mutually unbiased, if
\begin{equation}\label{eq15}
|\langle b_{1k}| b_{2i}\rangle|=\frac{1}{\sqrt{d}}, ~ \forall
k,i=1,2,\cdots,d.
\end{equation}
A set of orthonormal bases is said to be mutually unbiased if each
pair is mutually unbiased. In general, the maximal number of MUBs in
$d$ dimensions is an open problem. For a prime power $d$, one can
always construct a complete set of $d+1$ MUBs \cite{WK1,WK2,DT}.
MUBs was generalized by Kalev and Gour to MUMs in Ref. \cite{KA}. It
is shown that there always exists a complete set of $d+1$ MUMs for
arbitrary $d$. Two POVM measurements on $\mathcal{H}$,
$\mathcal{P}^{(b)}=\{P_{k}^{(b)}\}^d_{k=1}$, $b=1,2,$ are called
MUMs if
\begin{eqnarray}\label{eq16}
\mathrm{Tr}(P_{k}^{(b)})&=&1,
\nonumber\\\mathrm{Tr}(P_{k}^{(b)}P_{k'}^{(b')})&=&\frac{1}{d},~
b\neq b',\\
\mathrm{Tr}(P_{k}^{(b)}P_{k'}^{(b)})&=&\delta_{k,k'}\kappa+(1-\delta_{k,k'})\frac{1-\kappa}{d-1}
\nonumber,
\end{eqnarray}
where $\frac{1}{d}<\kappa\leq1$, and $\kappa=1$ if and only if all
$P_{k}^{(b)}$ are rank one projectors, i.e., $\mathcal{P}^{(1)}$ and
$\mathcal{P}^{(2)}$ are given by MUBs\cite{CB}.  Any complete set of
$d+1$ MUMs can be constructed as follows\cite{KA}. Let $\{F_{k,b}:
k=1,2,\cdots,d-1,b=1,2,\cdots,d+1\}$ be a set of $d^{2}-1$ traceless
Hermitian operators acting on $\mathcal{H}$ such that
$\mathrm{Tr}(F_{k,b}F_{k',b'})=\delta_{k,k'}\delta_{b,b'}$. Set
$F^{(b)}=\sum _{k=1}^{d-1} F_{k,b},b=1,2,\cdots,d+1$ and
\begin{eqnarray}\label{eq17}
\aligned
 F_k^{(b)}=
\begin{array}{cc}
\bigg\{
\begin{array}{cc}
F^{(b)}-(d+\sqrt{d}) F_{k,b} & k=1,2,\cdots ,d-1; \\
 (\sqrt{d}+1) F^{(b)} & k=d. \\
\end{array}
\\
\end{array}
\endaligned
\end{eqnarray}
Then $P_{k}^{(b)}=\frac{1}{d}I+tF_k^{(b)}$ with $k=1,2,\cdots,d,
b=1,2,\cdots,d+1$, which constitute a complete set of $d+1$ MUMs, as
long as $t$ is properly chosen such that all $P_{k}^{(b)}$ are
positive. The parameter
$\kappa=\frac{1}{d}+t^{2}(1+\sqrt{d})^2(d-1)$ is given by \cite{KA}.

With respect to a set of MUMs
$\mathcal{P}_{MUM}=\{\mathcal{P}^{(b)}\}^{d+1}_{b=1}$ the following
quantity has been defined \cite{CB2}:
$C({\rho},\mathcal{P}_{MUM})=\frac{1}{d+1}Q({\rho},\mathcal{P}_{MUM})=\frac{1}{d+1}\sum
_{b=1}^{d+1}Q({\rho},\mathcal{P}^{(b)})$, where
$Q({\rho},\mathcal{P}^{(b)})=\sum_{k=1}^{d}I_{\rho}(P_{k}^{(b)})$.
Base on the GWYD skew information, we define the following
generalized quantity:
$C^{\alpha,\beta}({\rho},\mathcal{P}_{MUM})=\frac{1}{d+1}Q^{\alpha,\beta}({\rho},
\mathcal{P}_{MUM})=\frac{1}{d+1}\sum
_{b=1}^{d+1}Q^{\alpha,\beta}({\rho},\mathcal{P}^{(b)})$, where
$Q^{\alpha,\beta}({\rho},\mathcal{P}^{(b)})=\sum
_{k=1}^{d}I_{\rho}^{\alpha,\beta}(P_{k}^{(b)})$. It is obvious that
$C^{\frac{1}{2},\frac{1}{2}}({\rho},\mathcal{P}_{MUM})=
C({\rho},\mathcal{P}_{MUM})$.

{\bf Theorem 1} With respect to MUMs
$\mathcal{P}_{MUM}=\{\mathcal{P}^{(b)}\}^{d+1}_{b=1}$,
$C^{\alpha,\beta}({\rho},\mathcal{P}_{MUM})$ satisfies the following
quantum uncertainty relations,
\begin{equation}\label{eq18}
C^{\alpha,\beta}({\rho},\mathcal{P}_{MUM})=\frac{\kappa
d-1}{(d^2-1)}Q^{\alpha,\beta}({\rho}),
\end{equation}
where $\alpha,\beta\geq 0$ and $\alpha+\beta\leq 1$.

 {\bf Proof.} Note that $\sum _{b=1}^{d+1}\sum
_{k=1}^{d}\mathrm{Tr}[(F_k^{(b)})^2\\ \rho]=(1+\sqrt{d})^2(d^2-1)$
\cite{CB}. Taking into account the relations $\sum _{b=1}^{d+1}\sum
_{k=1}^{d}\mathrm{Tr}(\rho^{\alpha}F_k^{(b)}\rho^{1-\alpha}F_k^{(b)})\\=(d+\sqrt{d})^2\sum
_{b=1}^{d+1}\sum_{k=1}^{d-1}\mathrm{Tr}(\rho^{\alpha}F_{k,b}\rho^{1-\alpha}F_{k,b})$
\cite{CB2} and $\sum _{b=1}^{d+1}\sum
_{k=1}^{d-1}(F_{k,b})^2=(d-\frac{1}{d})I$ \cite{SL5}, we have
\begin{align*}
&Q^{\alpha,\beta}({\rho},\mathcal{P}_{MUM})
\\&= \frac{1}{2}t^2\sum
_{b=1}^{d+1}\sum _{k=1}^{d}[\mathrm{Tr}[(F_k^{(b)})^2\rho]\\&\quad
+\mathrm{Tr}(\rho^{\alpha+\beta}F_k^{(b)}\rho^{1-\alpha-\beta}
F_k^{(b)})\\&\quad -\mathrm{Tr}(\rho^{\alpha}F_k^{(b)}
\rho^{1-\alpha}F_k^{(b)})-\mathrm{Tr}(\rho^{\beta}F_k^{(b)}
\rho^{1-\beta}F_k^{(b)})]
\\&=\frac{1}{2}t^2[(1+\sqrt{d})^2(d^2-1)\\&\quad+(d+\sqrt{d})^2(\sum
_{b=1}^{d+1}\sum
_{k=1}^{d-1}(\mathrm{Tr}(\rho^{\alpha+\beta}F_{k,b}\rho^{1-\alpha-\beta}F_{k,b})
\\&\quad-\mathrm{Tr}(\rho^{\alpha}F_{k,b}\rho^{1-\alpha}F_{k,b})
-\mathrm{Tr}(\rho^{\beta}F_{k,b}\rho^{1-\beta}F_{k,b})))]
\\&=\frac{1}{2}t^2[(1+\sqrt{d})^2(d^2-1)\\&\quad+(d+\sqrt{d})^2(\sum
_{b=1}^{d+1}\sum
_{k=1}^{d-1}(2I_{\rho}^{\alpha,\beta}(F_{k,b})-\mathrm{Tr}\rho{(F_{k,b})}^2)
\\&=\frac{\kappa
d-1}{(d-1)}Q^{\alpha,\beta}({\rho}).
\end{align*}
The theorem holds from the definition of
$C^{\alpha,\beta}({\rho},\mathcal{P}_{MUM})$. $\Box$

In particular, taking $\kappa=1$ in Theorem 1, we obtain the
following corollary.

{\bf Corollary 1} The quantum uncertainty relations based on the
generalized skew information with respect to MUB are given by
\begin{equation}\label{eq19}
C^{\alpha,\beta}({\rho},\mathcal{P}_{MUB})=\frac{1}{(d+1)}Q^{\alpha,\beta}({\rho}),
\end{equation}
where $\alpha,\beta\geq 0$ and $\alpha+\beta\leq 1$.

Corollary 1 can be viewed as a generalization of the corresponding
result in \cite{SL1}. Taking $\alpha=\beta=\frac{1}{2}$ in Theorem
1, we obtain the following corollary corresponding to the results
given in Ref. \cite{CB2}.

{\bf Corollary 2} The average coherence of a state $\rho$ with
respect to the $\mathcal{P}_{MUM}
=\{\mathcal{P}^{(b)}\}^{d+1}_{b=1}$ with parameter $\kappa$ is given
by
\begin{equation}\label{eq20}
C({\rho},\mathcal{P}_{MUM})=\frac{\kappa
d-1}{(d^2-1)}[d-(\mathrm{Tr}\sqrt{\rho})^2].
\end{equation}

By using Lemma 1, we can further prove the following theorem.

{\bf Theorem 2} The quantum complementarity relations based on
generalized skew information with respect to MUMs are given by
\begin{equation}\label{eq21}
C^{\alpha,\beta}({\rho},\mathcal{P}_{MUM})\leq\frac{\kappa
d-1}{2(d^2-1)}( d-\mathrm{Tr}{\rho^{\alpha}}
\mathrm{Tr}{\rho^{1-\alpha}})
\end{equation}
for $\alpha, \beta\in[0,1]$ with $\alpha+2\beta\leq1$ and
$2\alpha+\beta\leq1$.

Taking $\kappa=1$ in Theorem 2, we obtain the following corollary.

{\bf Corollary 3} The complementarity relations based on generalized
skew information with respect to MUBs are given by
\begin{equation}\label{eq22}
C^{\alpha,\beta}({\rho},\mathcal{P}_{MUB})\leq\frac{ 1}{2(d+1)}(
d-\mathrm{Tr}{\rho^{\alpha}} \mathrm{Tr}{\rho^{1-\alpha}})
\end{equation}
for $\alpha, \beta\in[0,1]$ with $\alpha+2\beta\leq1$ and
$2\alpha+\beta\leq1$.

\section{\bf  Uncertainty and complementarity relations based on
GWYD skew information with respect to general SIC-POVMs } In this
section, we study quantum uncertainty and complementarity relations
based on GWYD skew information with respect to general SIC-POVMs. A
set of $d^2$ positive-semidefinite operators $\{P_i\}_{i=1}^{d^2}$
is called a general SIC-POVM if
\begin{itemize}
\item $\sum _{i=1}^{d^2}P_i=\textbf{1}$, where $\textbf{1}$ is the identity matrix;
\item $\mathrm{Tr}{P_i}^2=a$ and $\mathrm{Tr}(P_k P_i)=\frac{1-
da}{d(d^2-1)}$, $\forall k,i\in\{1,2,\cdots,d^2\}$, $k\neq i$,
\end{itemize}
where $\frac{1}{d^3}<a\leq\frac{1}{d^2}$. $a=\frac{1}{d^2}$ if and
only if all $P_i$ are rank one, that is, the general SIC-POVM
becomes the SIC-POVM. Any general SIC-POVM can be constructed as
follows \cite{GG}. Let $\{F_i\}_{i=1}^{d^2-1}$ be a set of traceless
Hermitian operators on $\mathcal{H}$, satisfying
$\mathrm{Tr}(F_{i}F_{k})=\delta_{i,k}$. Set $F=\sum _{i=1}^{d^2-1}
F_{i}$. For any $t$ such that $P_i\geq0$ and
\begin{equation}\label{eq23}
a=\frac{1}{d^3}+t^2(d-1)(d+1)^3,
\end{equation}
one has
\begin{eqnarray*}P_i=
\begin{array}{cc}
 \bigg\{
\begin{array}{cc}
\frac{1}{d^2}I+t[F-d(d+1)F_{i}], i=1,\cdots,d^2-1; \\
\frac{1}{d^2}I+t(d+1)F,  i=d^2. \\
\end{array}
\\
\end{array}
\end{eqnarray*}

In Ref. \cite{CB2}, the coherence of a state with respect to a
general SIC-POVM $\{P_i\}_{i=1}^{d^2}$ with the parameter $a$ is
defined as $ C({\rho},\mathcal{P}_{GSM})=\sum
_{i=1}^{d^2}I_{\rho}(P_{i})$.  Now we define a generalized quantity
with respect to the GWYD skew information, $
C^{\alpha,\beta}({\rho},\mathcal{P}_{GSM})=\sum
_{i=1}^{d^2}I_{\rho}^{\alpha,\beta}(P_{i}), ~~~\alpha,\beta\geq 0,~
\alpha+\beta\leq 1$. It is straightforward to verify that
$C^{\frac{1}{2},\frac{1}{2}}({\rho},\mathcal{P}_{GSM})=C({\rho},\mathcal{P}_{GSM})$.

{\bf Theorem 3} The quantum uncertainty relations based on
generalized skew information with respect to a general SIC-POVM are
given by
\begin{equation}\label{eq24}
C^{\alpha,\beta}({\rho},\mathcal{P}_{GSM})=\frac{(ad^3-1)}{d(d^2-1)}Q^{\alpha,\beta}({\rho}),
\end{equation}
where $\alpha,\beta\geq 0$ and $\alpha+\beta\leq 1$.

{\bf Proof.}  Taking into account the relations $\sum
_{i=1}^{d^2}\mathrm{Tr}[(P_{i})^{2}\rho]=ad$ \cite{CB} and $\sum
_{i=1}^{d^2}\mathrm{Tr}(\rho^{\alpha}P_{i}\\
\rho^{1-\alpha}P_{i})=\frac{1}{d^2}+t^{2}d^{2}(d+1)^2\sum
_{i=1}^{d^2-1}\mathrm{Tr}(\rho^{\alpha}F_i\rho^{1-\alpha}\\F_i)$
\cite{CB2}, we have
\begin{align*}
&C^{\alpha,\beta}({\rho},\mathcal{P}_{GSM})
\\&=\sum
_{i=1}^{d^2}
\frac{1}{2}[\mathrm{Tr}[(P_i)^2\rho]+\mathrm{Tr}(\rho^{\alpha+\beta}P_i\rho^{1-\alpha-\beta}P_i)
\\&\quad-\mathrm{Tr}(\rho^{\alpha}P_i
\rho^{1-\alpha}P_i)-\mathrm{Tr}(\rho^{\beta}P_i \rho^{1-\beta}P_i)]
\\&=\frac{1}{2}[ad+(\frac{1}{d^2}+t^{2}d^{2}(d+1)^{2}(\sum
_{i=1}^{d^2-1}\mathrm{Tr}(\rho^{\alpha+\beta}F_i\rho^{1-\alpha-\beta}F_i)))
\\&\quad-(\frac{1}{d^2}+t^{2}d^{2}(d+1)^{2}(\sum
_{i=1}^{d^2-1}\mathrm{Tr}(\rho^{\alpha}F_i\rho^{1-\alpha}F_i)))\\&\quad-(\frac{1}{d^2}+t^{2}d^{2}(d+1)^{2}(\sum
_{i=1}^{d^2-1}\mathrm{Tr}(\rho^{\beta}F_i\rho^{1-\beta}F_i))]
\\&=\frac{1}{2}[ad-\frac{1}{d^2}+t^{2}d^{2}(d+1)^{2}(\sum
_{i=1}^{d^2-1}(\mathrm{Tr}(\rho^{\alpha+\beta}F_i\rho^{1-\alpha-\beta}F_i)
\\&\quad-\mathrm{Tr}(\rho^{\beta}F_i\rho^{1-\beta}F_i)-\mathrm{Tr}(\rho^{\alpha}F_i\rho^{1-\alpha}F_i))
\\&=\frac{(ad^3-1)}{d(d^2-1)}Q^{\alpha,\beta}({\rho}).
\end{align*}

Setting $a=\frac{1}{d^2}$ in Theorem 3, we obtain the following
corollary.

{\bf Corollary 4} The quantum uncertainty relations based on
generalized skew information with respect to a SIC-POVM are of the
form,
\begin{equation}\label{eq25}
C^{\alpha,\beta}({\rho}, \mathcal{P}_{SIC}
)=\frac{Q^{\alpha,\beta}({\rho})}{d(d+1)}
\end{equation}
for $\alpha,\beta\geq 0$ and $\alpha+\beta\leq 1$.

In particular, taking $\alpha=\beta=\frac{1}{2}$, we obtain the
following corollary corresponding to result in Ref. \cite{CB2}.

{\bf Corollary 5} The coherence with respect to a general SIC-POVM
is given by
\begin{equation}\label{eq26}
C({\rho}, \mathcal{P}_{GSM}
)=\frac{(ad^3-1)(d-(\mathrm{Tr}\sqrt{\rho})^2)}{d(d^2-1)}.
\end{equation}

By using Lemma 1, we can prove the following theorem.

{\bf Theorem 4} The quantum complementarity relations based on
generalized skew information with respect to a general SIC-POVM are
given by
\begin{equation}\label{eq27}
C^{\alpha,\beta}({\rho},\mathcal{P}_{GSM})\leq\frac{(ad^3-1)}{2d(d^2-1)}(
d-\mathrm{Tr}{\rho^{\alpha}} \mathrm{Tr}{\rho^{1-\alpha}})
\end{equation}
for $\alpha, \beta\in[0,1]$ with $\alpha+2\beta\leq1$ and
$2\alpha+\beta\leq1$.

Taking $a=\frac{1}{d^2}$ in Theorem 4, we obtain the following
corollary.

{\bf Corollary 6} The complementarity relations based on generalized
skew information with respect to a SIC-POVM are given by
\begin{equation}\label{eq28}
C^{\alpha,\beta}({\rho},\mathcal{P}_{SIC})\leq\frac{1}{2d(d+1)}(
d-\mathrm{Tr}{\rho^{\alpha}}
\mathrm{Tr}{\rho^{1-\alpha}})\end{equation} for $\alpha,
\beta\in[0,1]$ with $\alpha+2\beta\leq1$ and $2\alpha+\beta\leq1$.

{\bf Example 1} Consider the Werner state,
$$\rho_w=\left(\begin{array}{cccc}
         \frac{1}{3}p&0&0&0\\
         0&\frac{1}{6}(3-2p)&\frac{1}{6}(4p-3)&0\\
         0&\frac{1}{6}(4p-3)&\frac{1}{6}(3-2p)&0\\
         0&0&0&\frac{1}{3}p\\
         \end{array}
         \right),
$$
where $p\in [0,1]$. $\rho_w$ is separable when $p\in
[0,\frac{1}{3}]$. Take $ \kappa=1$ and $a=\frac{1}{d^2}$, Figures 1
and 2 illustrate the complementarity relations of Eqs.~(\ref{eq22})
and (\ref{eq28}) with different values of $\alpha$ and $\beta$,
respectively.

\begin{figure}[htbp]\centering
{\begin{minipage}[b]{0.8\linewidth}
\includegraphics[width=1\textwidth]{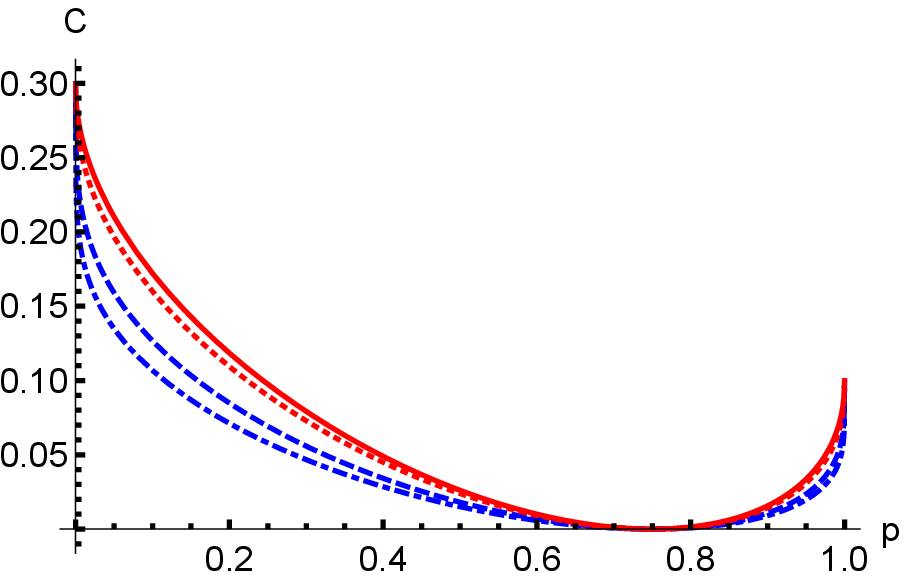}
\end{minipage}}
\caption{The $C$-axis shows the complementarity and its upper
bounds. Red solid (dotted) line represents the value of the right
hand side of Eq.~(\ref{eq22}) with $\alpha=\frac{5}{12}$
($\alpha=\frac{1}{3}$) and $\beta=\frac{1}{6}$ ($\beta=\frac{1}{4}$)
for $\rho_w$; blue dashed (dotdashed) line represents the value of
the left hand side of Eq.~(\ref{eq22}) with $\alpha=\frac{1}{3}$
($\alpha=\frac{5}{12}$) and $\beta=\frac{1}{4}$
($\beta=\frac{1}{6}$) for $\rho_w$.} \label{fig:u3}
\end{figure}

\begin{figure}[htbp]\centering
{\begin{minipage}[b]{0.8\linewidth}
\includegraphics[width=1\textwidth]{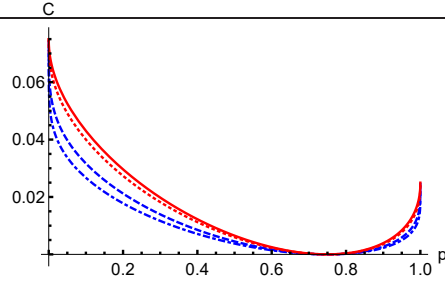}
\end{minipage}}
\caption{The $C$-axis shows the complementarity and its upper
bounds. Red solid (dotted) line represents the value of the right
hand side of Eq.~(\ref{eq28}) with $\alpha=\frac{5}{12}$
($\alpha=\frac{1}{3}$) and $\beta=\frac{1}{6}$ ($\beta=\frac{1}{4}$)
for $\rho_w$; blue dashed (dotdashed) line represents the value of
the left hand side of Eq.~(\ref{eq28}) with $\alpha=\frac{5}{12}$
($\alpha=\frac{1}{3}$) and $\beta=\frac{1}{6}$ ($\beta=\frac{1}{4}$)
for $\rho_w$.} \label{fig:u5}
\end{figure}

\section{\bf  Conclusions}
Based on GWYD skew information, we have derived the uncertainty and
complementarity relations with respect to MUMs and general
SIC-POVMs, which include some uncertainty relations and
complementarity relations in \cite{SL2} and \cite{CB2} as special
cases. It is worth noting that the uncertainty and complementarity
relations we obtained are all state-dependent. Our approaches and
results may shed some new light on further investigations on quantum
coherence and complementary measurements.

\acknowledgments This work was supported by National Natural Science
Foundation of China (Grant Nos. 11701259, 11461045, 11771198,
11675113); Jiangxi Provincial Natural Science Foundation (Grant No.
20202BAB201001); Beijing Municipal Commission of Education
(KZ201810028042); Beijing Natural Science Foundation (Grant No.
Z190005); Academy for Multidisciplinary Studies, Capital Normal
University; Shenzhen Institute for Quantum Science and Engineering,
Southern University of Science and Technology, Shenzhen 518055,
China (No. SIQSE202001).

\end{document}